\documentclass[superscriptaddress,prd,aps,showpacs,nofootinbib,showkeywords,eqsecnum,preprint]{revtex4-1}

\usepackage{graphicx,color,amsmath,amsxtra}
\usepackage{epsf}
\usepackage{amssymb}
\usepackage{enumerate}
\usepackage{hhline}
\usepackage{array}
\usepackage{tabularx}
\usepackage[unicode]{hyperref}
\usepackage{graphicx}                
\usepackage{epstopdf}

\begin{document}
\pagestyle{myheadings}
\title{Stability analysis of de Sitter solution in the Einstein-Grisaru-Zanon gravity using the dynamical system method}
\author{Tuan Q. Do}
\email{tuan.doquoc@phenikaa-uni.edu.vn}
\affiliation{Phenikaa Institute for Advanced Study, Phenikaa University, Hanoi 12116, Vietnam}

\begin{abstract}
In this paper, we would like to investigate the stability of de Sitter solution in the Einstein-Grisaru-Zanon gravity, which is a novel fourth-order gravity model considered recently in a paper [Phys. Lett. B {\bf 855} (2024) 138811]. As a result, we are able to derive the corresponding field equations for the Einstein-Grisaru-Zanon gravity by using an effective method based on the Euler-Lagrange equations. Unfortunately, one of the obtained field equations does not coincide with that derived in the original paper of the Einstein-Grisaru-Zanon gravity due to a gap between higher-order derivative terms.  However, our de Sitter solution is still identical to one solved in the original paper of the Einstein-Grisaru-Zanon gravity due to the vanishing of the gap.  Furthermore, a stability analysis based on the dynamical system method is performed to indicate that the obtained de Sitter solution is always unstable, no matter it presents an inflationary phase or expanding phase of universe. This result confirms the validity of stability investigation carried out in the original paper of the Einstein-Grisaru-Zanon gravity. 
\end{abstract}
\maketitle
\section{Introduction} \label{intro}
Higher-order gravities, in particularly fourth-order gravities or quadratic gravities, have been one of the most promising generalizations of Einstein's general relativity in order to resolve the remaining cosmological issues, which cannot be explained within the Einstein's gravity \cite{Schmidt:2006jt,Alvarez-Gaume:2015rwa,Salvio:2018crh}. Two typical issues can be listed here are the cosmic inflation of early universe \cite{Starobinsky:1980te,Whitt:1984pd,Maeda:1987xf,Barrow:1988xh} and the accelerated expansion of late-time universe \cite{Carroll:2004de,Copeland:2006wr,Nojiri:2010wj,Nojiri:2017ncd}. Besides, it is expected that a renormalizable model of gravity may be constructed if higher-order curvature correction terms, e.g., $R^2$ and $R_{\mu\nu}R^{\mu\nu}$, were introduced into the pure Einstein-Hilbert action \cite{Stelle:1976gc}.  Among the proposed higher-order gravities, the Starobinsky model \cite{Starobinsky:1980te} has been shown to be highly consistent with the recent cosmic microwave background (CMB)  radiation probes such as the Planck satellite \cite{Akrami:2018odb}.  It turns out that the success of the Starobinsky model is due to the contribution of the Ricci scalar squared term, i.e, $R^2$, which is nothing but a quantum correction \cite{Starobinsky:1980te}. An additional interesting point of the Starobinsky model is that it is a special fourth-order gravity model being free from the well-known  Ostrogradsky ghost \cite{Woodard:2015zca}. However, the more precise CMB measurements, e.g., the very recent data of the Atacama Cosmology Telescope (ACT) \cite{ACT:2025tim}, may address generalizations of the Starobinsky model, e.g., see Refs. \cite{Myrzakulov:2014hca,Cano:2020oaa,Rodrigues-da-Silva:2021jab,Ivanov:2021chn,Modak:2022gol,Ketov:2022lhx,Ketov:2022zhp,Do:2023yvg,Pham:2024fub,Asorey:2024oxw} for the recent ones. On the other hand, searching other fourth-order gravities instead of the Starobinsky one has been another interesting topic. For example, some novel fourth-order gravities such as the so-called Einsteinian cubic gravity \cite{Cano:2020oaa,Bueno:2016xff,Arciniega:2018fxj}, Ricci-inverse gravity \cite{Amendola:2020qho,Do:2020vdc,Do:2021fal}, Einstein-Bel-Robinson  gravity \cite{Ketov:2022lhx,Ketov:2022zhp,Do:2023yvg,Sajadi:2023bwe}, and Einstein-Grisaru-Zanon gravity \cite{CamposDelgado:2024jst,Toyama:2024ugg,Mustafa:2025djp} have been proposed recently. 

In harmony with our previous investigations of the stability of de Sitter inflationary solutions within some novel fourth-order gravity models \cite{Do:2023yvg,Pham:2024fub,Do:2020vdc,Do:2021fal}, we would like to revisit this issue in the Einstein-Grisaru-Zanon (EGZ) gravity, which has been studied recently in Ref. \cite{CamposDelgado:2024jst} (see Ref. \cite{Toyama:2024ugg} for its Starobinsky-Grisaru-Zanon extension and Ref. \cite{Mustafa:2025djp} for related works on black hole physics). It is apparent  that the EGZ gravity is a non-trivial extension of Einstein's general relativity, which has been constructed by introducing the leading superstring correction term firstly derived by Grisaru and Zanon in Ref. \cite{Grisaru:1986vi} into the pure Einstein-Hilbert action. Remarkably, a de Sitter solution has been found in the EGZ gravity due to the contribution of the Grisaru-Zanon (GZ) term. Additionally, this de Sitter solution has been shown to be generically unstable \cite{CamposDelgado:2024jst}. According to the discussions in Refs. \cite{Elizalde:2014xva,Pozdeeva:2019agu,Vernov:2021hxo}, the instability of de Sitter solution plays an important criteria for claiming that the EGZ gravity could potentially be a realistic inflationary model since it would not face the so-called eternal inflation issue \cite{Guth:2007ng}, which can lead to a multiverse scenario. In this case, a quasi-de Sitter inflationary solution is expected to exist. This view is strongly motivated by an interesting fact that the Starobinsky model does not admit an exact de Sitter solution but a quasi-de Sitter one \cite{Starobinsky:1980te}. Examining the existence and/or stability of de Sitter solution in higher-order gravity models should be one of the most important initial checks before investigating further their cosmological viability and consequences, e.g., see Refs. \cite{Elizalde:2014xva,Pozdeeva:2019agu,Vernov:2021hxo,Faraoni:2004dn,Faraoni:2005vk,Toporensky:2006kc,Kamenshchik:2024kay} for examples.

According to our obtained results, which will be presented in the following sections, we would like to highlight here two main results of our present paper: (i) The value of our obtained de Sitter solution is identical to one derived in the original paper of the EGZ gravity \cite{CamposDelgado:2024jst} despite an important issue that one field equation, which is derived by using an effective method based on the Euler-Lagrange (EL) equations, is not identical to that obtained in Ref. \cite{CamposDelgado:2024jst}  and (ii) our stability analysis based on the dynamical system method concludes that the obtained de Sitter solution is generically unstable, consistent with the finding made in Ref. \cite{CamposDelgado:2024jst}. Additionally, this de Sitter solution will be numerically verified as a repeller.

In summary, this paper will be organized as follows: (i) Its brief introduction has been written in Sec. \ref{intro}. (ii) Basic setup of the EGZ gravity involving its action, its field equations, and its de Sitter solution will be presented in Sec. \ref{sec2}.  (iii) Stability of the obtained de Sitter solution will be investigated carefully through the dynamical system method in Sec. \ref{sec3}. (iv) Finally, main results of the present paper will be concluded in Sec. \ref{final}.  
\section{Basic setup} \label{sec2}
\subsection{Einstein-Grisaru-Zanon gravity}
An action of the EGZ gravity considered in Ref. \cite{CamposDelgado:2024jst} is given by
\begin{equation}
S_{\rm EGZ}= \frac{M_p^2}{2}\int d^4 x \sqrt{-g}  \left(R+ \frac{\gamma}{M_p^6} J \right),
\end{equation}
where $M_p\equiv 1/\sqrt{8\pi G}$ is the reduced Planck mass, $\gamma$ is a dimensionless coupling constant, and $J$ is nothing but the leading superstring correction firstly derived in Ref. \cite{Grisaru:1986vi} by Grisaru and Zanon,
\begin{equation}
J = \left( R^{\mu\rho\sigma\nu} R_{\lambda\rho\sigma\tau} +\frac{1}{2} R^{\mu\nu\rho\sigma}R_{\lambda \tau \rho\sigma}\right)R_\mu{}^{\alpha \beta \lambda}R^\tau{}_{\alpha\beta\nu}.
\end{equation}
 It should noted that the appearance of $M_p^{-6}$ in front of $J$ in the above action is due to the result that $J$ is nothing but the $\alpha'^3$ superstring correction as pointed out in Ref. \cite{Grisaru:1986vi}. Furthermore, it is well known that $\alpha'$ stands for an inverse string tension with $\alpha' \sim M_p^{-2}$. Therefore, one can write $\gamma M_p^{-6}$ instead of $\alpha'^3$, where $\gamma$ is an additional constant. It has been argued in Ref. \cite{CamposDelgado:2024jst} that the value of $\gamma$ depends upon a compactification from ten to four dimensions and the unknown vacuum expectation value of string dilaton. In addition,  an upper bound on $\gamma$ such as $\gamma < 1.62 \times 10^{-5}$ has been worked out in Ref. \cite{CamposDelgado:2024jst} from the Hawking temperature. 

Very interestingly, it has been shown in Ref. \cite{CamposDelgado:2024jst} that the EGZ gravity admits a Schwarzschild black hole solution, which will be identical to that derived from the Einstein-Bel-Robinson (EBR) gravity, whose action is given by (without the $R^2$-term)
\cite{Ketov:2022lhx,Ketov:2022zhp,Do:2023yvg,Sajadi:2023bwe}
\begin{equation}
S_{\rm EBR} =  \frac{M_p^2}{2}\int d^4 x \sqrt{-g}  \left(R -\frac{\beta}{32M_p^6}T^2 \right),
\end{equation}
if $\gamma =\beta$. In this action, $T^2$ is nothing but the Bel-Robinson (BR) tensor squared defined as $T^2 \equiv T_{\rho \nu \lambda \mu}T^{\rho \nu \lambda \mu}$ with \cite{Bel:1959uwe,Robinson:1959ev,Deser:1999jw}
 \begin{equation}
 T^{\mu\nu\lambda\rho} =R^{\mu\alpha\beta\lambda} R^\nu{}_{\alpha\beta}{}^\rho +R^{\mu\alpha\beta\rho}R^\nu{}_{\alpha\beta}{}^\lambda-\frac{1}{2}g^{\mu\nu}R^{\alpha\beta\gamma\lambda}R_{\alpha\beta\gamma}{}^\rho.
 \end{equation}
 
 However, a significant gap between these two EGZ and EBR gravities will appear in the spatially flat Friedmann-Lemaitre-Robertson-Walker (FLRW) background as pointed out in Ref. \cite{CamposDelgado:2024jst}. Similar things could also appear in other background spacetimes. It is noted that a de Sitter solution has been shown to exist within the EGZ gravity \cite{CamposDelgado:2024jst}. Furthermore, it turns out to be unstable \cite{CamposDelgado:2024jst}. This result indicates that the EGZ gravity could be relevant to the inflationary phase of our early universe, since it would not lead to the so-called eternal inflation along with the associated multiverse scenario \cite{Guth:2007ng}, according to discussions in Refs. \cite{Elizalde:2014xva,Pozdeeva:2019agu,Vernov:2021hxo}. In this paper, we would like to investigate the stability of de Sitter solution of the EGZ gravity using the dynamical system, which seems to be the powerful method for studying the stability of cosmological solutions \cite{Bahamonde:2017ize}. Main results will be presented in the rest of paper. 
\subsection{Field equations}
Firstly, we consider the following spatially flat FLRW metric given by
\begin{equation} \label{metric}
ds^2 =-N^2(t)dt^2 +e^{2\alpha(t)} \left(dx^2 +dy^2 +dz^2 \right),
\end{equation}
where  $N(t)$ is the lapse function of cosmic time $t$, which is introduced to derive the corresponding Friedmann equation from its EL equation \cite{Myrzakulov:2014hca,Asorey:2024oxw,Do:2023yvg,Pham:2024fub,Do:2020vdc,Toporensky:2006kc,Kao:1991zz} and $\alpha(t)$ is a scale factor encoding the evolution of universe. It is known that after deriving its corresponding Friedmann equation along with another field equation of the scale factor $\alpha(t)$, the lapse function $N(t)$ should be set to be one \cite{Myrzakulov:2014hca,Asorey:2024oxw,Do:2023yvg,Pham:2024fub,Do:2020vdc,Toporensky:2006kc,Kao:1991zz}.  

To derive the corresponding field equations of the EGZ gravity for the FLRW metric, we prefer to use an effective calculation approach based on the EL equations, which have been used not only by us \cite{Do:2023yvg,Pham:2024fub,Do:2020vdc,Do:2021fal} but also by other people, e.g., see Ref. \cite{Myrzakulov:2014hca} and especially Ref. \cite{Asorey:2024oxw}. To do this, we need to define an explicit expression of the EGZ Lagrangian, whose definition is given by \cite{xAct}
 \begin{equation}
 {\cal L} =\sqrt{-g} \left(R+\bar\gamma J \right),
 \end{equation}
 where  $\bar\gamma \equiv \gamma/M_p^6$ as an additional parameter introduced for convenience.
For the FLRW metric, it appears that \cite{Do:2023yvg,Pham:2024fub}
\begin{align}
 \sqrt{-g} & =N e^{3\alpha}, \\
  R & = -6N^{-2} \left(N^{-1} \dot N \dot\alpha -\ddot\alpha-2\dot\alpha^2 \right),
 \end{align}
 where $\dot\alpha \equiv d\alpha/dt$ as well as $\ddot\alpha \equiv d^2\alpha/dt^2$.  As a result, the corresponding value of $J$ is defined to be \cite{xAct}
 \begin{equation}
J = \frac{6}{N^{12}}\left(j_4 N^4 + j_3N^3 +j_2N^2 +j_1 N+j_0 \right),
\end{equation}
with
\begin{align}
j_4&= \ddot\alpha^4+8\dot\alpha^2 \ddot\alpha^3 +22 \dot\alpha^4 \ddot\alpha^2 + 24\dot\alpha^6 \ddot\alpha+12\dot\alpha^8,\\
j_3&= -4\dot N \dot\alpha \left(\ddot\alpha^3 +6\dot\alpha^2 \ddot\alpha^2 +11 \dot\alpha^4 \ddot\alpha +6\dot\alpha^6 \right),\\
j_2&= 2\dot N^2 \dot\alpha^2 \left( 3\ddot\alpha^2 +12 \dot\alpha^2 \ddot\alpha +11\dot\alpha^4 \right),\\
j_1&= -4\dot N^3 \dot\alpha^3 \left(\ddot\alpha +2\dot\alpha^2 \right),\\
j_0 &= \dot N^4 \dot\alpha^4.
\end{align}
 It is apparent  that ${\cal L}$ is a functional of the second-order time derivative of $\alpha(t)$ and the first-order time derivative of $N(t)$. Therefore, its EL equation for $N(t)$ takes the following form,
\begin{align}
\frac{\partial {\cal L}}{\partial N} -\frac{d}{dt} \left(\frac{\partial {\cal L}}{\partial \dot N}\right)=0,
\end{align}
which turns out to be
\begin{align} \label{equation-1-FLRW}
\dot \alpha ^2 + \bar\gamma \left[ 4 \dot\alpha \left(3\ddot\alpha^2 + 12 \dot\alpha^2 \ddot\alpha +11 \dot\alpha^4 \right)  \alpha^{(3)}  - 3 \ddot\alpha^4 + 28 \dot\alpha^2 \ddot\alpha^3 + 138 \dot\alpha^4 \ddot\alpha^2 + 132 \dot\alpha^6 \ddot\alpha -12  \dot\alpha^8 \right]  =0,
\end{align}
after setting $N(t)=1$.
On the other hand,  its  EL equation for $\alpha(t)$ reads
\begin{align}
\frac{\partial {\cal L}}{\partial \alpha} -\frac{d}{dt} \left(\frac{\partial {\cal L}}{\partial \dot \alpha}\right) + \frac{d^2}{dt^2} \left(\frac{\partial {\cal L}}{\partial \ddot\alpha}\right)&=0,
\end{align}
which becomes
\begin{align}\label{equation-2-FLRW}
&2\ddot\alpha+ 3\dot\alpha^2 + \bar\gamma \left[ 4 \left(3\ddot\alpha^2 + 12 \dot\alpha^2 \ddot\alpha +11 \dot\alpha^4 \right) \alpha^{(4)} + 24 \left( \ddot\alpha +2 \dot\alpha^2 \right) \left(\alpha^{(3)} \right)^2  \right.\nonumber\\
&\left. +  \dot\alpha \left( 264 \ddot\alpha^2 +640 \dot\alpha^2 \ddot\alpha +264 \dot\alpha^4 \right)  \alpha^{(3)} +47 \ddot \alpha^4 +636 \dot\alpha^2 \ddot\alpha^3 +1206 \dot\alpha^4 \ddot\alpha^2 + 300 \dot\alpha^6 \ddot\alpha  -36 \dot\alpha^8  \right] =0,
\end{align}
also after setting $N(t)=1$. Here, $\alpha^{(n)} \equiv d^n \alpha(t)/dt^n$ is nothing but the $n$th-order time derivative of $\alpha(t)$. 
It is noted that these two field equations can be rewritten in terms of the Hubble parameter, which is defined as $H(t)\equiv \dot a(t)/a(t) = \dot\alpha(t)$.  Furthermore, they encode the dynamics of the EGZ gravity in the FLRW background spacetime.  Additionally, the EGZ gravity can be called as a fourth-order gravity model due to the obtained result that its field equation \eqref{equation-2-FLRW} is nothing but a fourth-order ordinary differential  equation (ODE), while its Friedmann equation \eqref{equation-1-FLRW} is a third-order ODE. A straightforward comparison can be made to confirm that Eq. \eqref{equation-1-FLRW} is identical to Eq. (23) in Ref. \cite{CamposDelgado:2024jst}, while Eq. \eqref{equation-2-FLRW} is different from Eq. (24) in Ref. \cite{CamposDelgado:2024jst}, provided that $H =\dot\alpha$, $\dot H =\ddot\alpha$, $\ddot H = \alpha^{(3)}$, and $H^{(3)} = \alpha^{(4)}$. The gap between Eq. \eqref{equation-2-FLRW} and Eq. (24) of Ref. \cite{CamposDelgado:2024jst} is due to the difference in  their higher-order derivative terms. Very interestingly,  this gap will disappear into a de Sitter solution as shown below.
\subsection{de Sitter solution}
Given the field equations of the EGZ gravity, we now would like to seek an exact de Sitter solution, with the hint from the previous investigations on fourth-order gravities \cite{Do:2023yvg,Pham:2024fub,Do:2020vdc,Toporensky:2006kc,Barrow:2005qv,Barrow:2006xb,Barrow:2009gx}, by taking an ansatz for the scale factor, 
\begin{equation}
\alpha(t) =\zeta t,
\end{equation}
here $\zeta$ is an unknown parameter. 
As a result, both field equations \eqref{equation-1-FLRW} and \eqref{equation-2-FLRW} lead to the same equation of $\zeta$,
\begin{equation} \label{equation-of-zeta-iso}
12 \bar \gamma \zeta^6 -1 =0,
\end{equation}
which is easily solved to give us a non-trivial exact solution of $\zeta$,
\begin{equation} \label{sol-of-zeta-FLRW}
\zeta =\left( \frac{1}{12\bar\gamma } \right)^{\frac{1}{6}}.
\end{equation}
This  de Sitter solution is indeed identical to that found in Ref. \cite{CamposDelgado:2024jst}.
It is apparent that $0<\bar\gamma \ll 1$ is an efficient constraint to obtain the corresponding de Sitter inflationary solution with  $\zeta \gg 1$. On the other hand, the corresponding de Sitter expanding solution with $0<\zeta \leq 1$ will exist for $\bar\gamma \geq 1/12$. Of course, any negative value of $\bar\gamma$ will not be suitable for the existence of de Sitter expanding or inflationary solutions. 
\section{Dynamical system} \label{sec3}
\subsection{Autonomous equations}
In this section, we would like to address an important question of whether the obtained de Sitter solution is stable or not by using the dynamical system method. To do this, we will transform the field equations into the corresponding dynamical system, similar to the previous works on fourth-order gravity models in Refs. \cite{Do:2023yvg,Pham:2024fub,Do:2020vdc,Toporensky:2006kc,Barrow:2005qv,Barrow:2006xb,Barrow:2009gx}, by introducing dimensionless dynamical variables such as
 \begin{align}
& B=\frac{1}{\dot\alpha^2},\\
&Q=\frac{\ddot\alpha}{\dot\alpha^2},\\
& Q_2 =\frac{\alpha^{(3)}}{\dot\alpha^3}, 
 \end{align}
 here the Hubble parameter is given by $H=\dot\alpha$. Note that the notations of dynamical variables have been taken from Refs. \cite{Barrow:2005qv,Barrow:2006xb,Barrow:2009gx}.  As a result, the corresponding set of autonomous equations of dynamical variables is given by 
 \begin{align}
  \label{Dyn-1}
 B' &= -2QB,\\
  \label{Dyn-2}
 Q' &=Q_2 -2Q^2,\\
 \label{Dyn-3}
 Q_2 '&= \frac{\alpha^{(4)}}{\dot\alpha^4}-3Q Q_2,
 \end{align}
 where $'$ should be understood as a derivative w.r.t. a dynamical time variable $\tau$ defined as $\tau =\int \dot\alpha dt$. To be more specific, $B' =dB/d\tau$ and so on. In Eq. \eqref{Dyn-3}, there is an undetermined term ${\alpha^{(4)}}/{\dot\alpha^4}$, which can be determined from the field equation \eqref{equation-2-FLRW}. In particular, Eq. \eqref{equation-2-FLRW} can be rewritten in terms of the dynamical variables  as follows
 \begin{align} \label{Dyn-4}
 &2B^3 Q+3B^3 +\bar\gamma \left[ 4 \left(3Q^2 +12 Q +11  \right) \frac{\alpha^{(4)}}{\dot\alpha^4} +24 \left( Q+ 2 \right)  Q_2^2 + \left(264 Q^2 +640 Q +264  \right) Q_2 \right. \nonumber\\
 & \left. + 47 Q^4 +636  Q^3 +1206  Q^2 + 300 Q -36 \right] =0.
\end{align}
A straightforward calculation can be made to have
\begin{align} \label{Dyn-4-1}
\frac{\alpha^{(4)}}{\dot\alpha^4} = & \frac{-1}{4\bar\gamma \left(3Q^2 +12 Q +11  \right) } \left\{ 2B^3 Q+3B^3 + \bar\gamma \left[24 \left( Q+ 2 \right)  Q_2^2 + \left(264 Q^2 +640 Q +264  \right) Q_2 \right. \right. \nonumber\\
&\left.\left. + 47 Q^4 +636  Q^3 +1206  Q^2 + 300 Q -36  \right] \right\}.
\end{align}
 Besides, the Friedmann equation \eqref{equation-1-FLRW} can be expressed  in terms of the dynamical variables such as
 \begin{align} \label{Dyn-5}
 B^3 +\bar \gamma \left[ 4\left( 3Q^2 +12 Q +11  \right) Q_2 - 3 Q^4 + 28 Q^3 +138 Q^2 + 132 Q -12  \right] =0,
 \end{align}
 which any found fixed points must satisfy fully. 
 
  In conclusion, we have derived the corresponding dynamical system of the EGZ gravity, which is formed by the autonomous equations \eqref{Dyn-1}, \eqref{Dyn-2}, and \eqref{Dyn-3} along with ${\alpha^{(4)}}/{\dot\alpha^4}$ defined in Eq. \eqref{Dyn-4-1} and the constraint equation \eqref{Dyn-5}. In principle, the key relationship between the dynamical system and field equations in the FLRW background is based on the term ${\alpha^{(4)}}/{\dot\alpha^4}$, whose explicit expression can only be determined from the fourth-order field equation \eqref{equation-2-FLRW}. On the other hand, the dynamical variables will be tightly constrained by the third-order field equation \eqref{equation-1-FLRW}. All these things indicate that any constraint of field parameters in the field equations will be met exactly in the dynamical system. For example, the positivity constraint of $\bar\gamma$  due to the existence of the de Sitter solution will be re-casted in the next subsection, where the existence of fixed point of the dynamical system will be investigated.
 \subsection{Fixed points}
 Given the above dynamical system, we are going to seek its (isotropic) fixed points following the previous papers \cite{Do:2023yvg,Pham:2024fub,Do:2020vdc,Toporensky:2006kc,Barrow:2005qv,Barrow:2006xb,Barrow:2009gx}. In particular, fixed points obtained in these papers can be shown to be equivalent to particular cosmological solutions of field equations such as de Sitter solutions. Therefore, the stability of these fixed points will tell us the stability of the corresponding cosmological solutions. This strategy turns out to be very useful when field equations are complicated due to the existence of higher-order terms. As shown above, the dynamical system involves only the first-order ODEs and  is therefore less complicated. For more discussions of the usefulness of the dynamical system as well as its fixed points, see a very interesting review paper \cite{Bahamonde:2017ize}.
 
  Mathematically, fixed points are solutions of the following set of equations \cite{Do:2023yvg,Pham:2024fub,Do:2020vdc,Toporensky:2006kc,Bahamonde:2017ize,Barrow:2005qv,Barrow:2006xb,Barrow:2009gx}, 
 \begin{equation}
B'=Q'=Q_2' =0.
\end{equation}
Consequently, we have for $B \neq 0$ 
 \begin{equation}
 Q=Q_2=\frac{\alpha^{(4)}}{\dot\alpha^4}=0.
 \end{equation}
 As a result, both equations \eqref{Dyn-4} and   \eqref{Dyn-5} reduce to
 \begin{equation}
 B^3 -12\bar \gamma  =0.
 \end{equation}
  Since $B>0$ then $\bar\gamma >0$ is required accordingly. Interestingly, this constraint is identical to that of the obtained de Sitter solution.
 Integrating this equation leads to an exact solution (up to an integration constant),
 \begin{equation}
 \alpha(t)= \zeta t,
 \end{equation}
with $\zeta$ being defined in Eq. \eqref{sol-of-zeta-FLRW}. This result  confirms our expectation that the found (isotropic) fixed point is indeed equivalent to the de Sitter solution found in the previous section. It is noted that there exists another fixed point to the dynamical system with $B=0$, according to Eq. \eqref{Dyn-1}. As a consequence, the corresponding equation of $Q$ can be defined to be
\begin{equation}
21Q^4 +124Q^3 +226Q^2 +132Q-12=0.
\end{equation}
Solving this equation gives us two real roots, $Q \simeq -3.1199$ and $Q\simeq 0.0796$.
However, this fixed point is not equivalent to the de Sitter solution and  is therefore not our desired one. For convenience, we call it the non-de Sitter fixed point.
 \subsection{Stability of the de Sitter fixed point}
We have found the fixed point, which is indeed equivalent to the de Sitter solution of Einstein field equations.  Therefore, instead of investigating the stability of the de Sitter solution we will consider the stability of this fixed point. For convenience, we will call it the de Sitter fixed point. We would like to remind that the instability of the de Sitter solution of the EGZ gravity has been verified by a different simple approach in Ref. \cite{CamposDelgado:2024jst}.

\begin{figure}\centering
	  \includegraphics[scale=0.5]{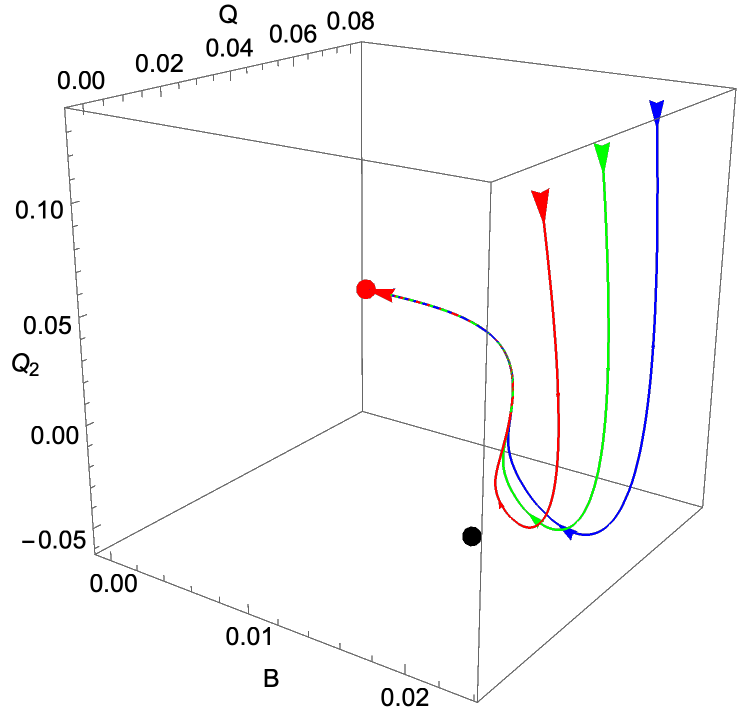}
	
\caption{\it The fixed point, which is equivalent to the de Sitter solution and is  displayed as a black point with $(B,Q,Q_2) \simeq (0.023,0,0)$, acts as a repeller since  all three trajectories tend to repel it. Instead, they tend to converge to the non-de Sitter fixed point displayed as a red point with $(B,Q,Q_2) \simeq (0,0.0796,0.0127)$. Different colors correspond to different initial conditions. The parameters have been chosen as $\gamma=10^{-6}$ and $M_p=1$.}
\label{fig1}
\end{figure}

As a standard procedure, we first perturb the autonomous equations \eqref{Dyn-1}, \eqref{Dyn-2}, and \eqref{Dyn-3} around the de Sitter fixed point,
\begin{align}
\label{pert-1}
\delta B' &=-2B\delta Q,\\
\label{pert-2}
\delta Q' &= \delta Q_2,\\
\label{pert-3}
\delta Q_2' &=- \frac{1}{44 \bar\gamma} \left[ 2B^3 \delta Q + 9B^2 \delta B  + 264\bar\gamma \delta Q_2 +300\bar\gamma \delta Q \right],
\end{align}
with the help of the result,
\begin{equation} \label{pert-4}
\delta \left(\frac{\alpha^{(4)}}{\dot\alpha^4} \right) =- \frac{1}{44\bar\gamma} \left( 2B^3 \delta Q + 9B^2 \delta B  + 264\bar\gamma \delta Q_2 +300 \bar\gamma \delta Q \right),
\end{equation}
which is derived from Eq. \eqref{Dyn-4-1}. Furthermore, taking perturbations from Eq. \eqref{Dyn-5}, we can simplify Eq. \eqref{pert-3} as follows
\begin{equation}\label{pert-5}
\delta Q_2' = -\frac{1}{44\bar\gamma} \left( 2B^3 \delta Q -96\bar\gamma \delta Q  + 132 \bar\gamma \delta Q_2  \right).
\end{equation}
By taking exponential perturbations, 
\begin{align}
\delta B &= C_B \exp[\mu\tau], \\
\delta Q &= C_Q \exp[\mu\tau],\\
\delta Q_2 &= C_{Q_2}  \exp[\mu\tau],
\end{align}
we are able to obtain the following homogeneous system of linear equations   coming from the perturbed equations \eqref{pert-1}, \eqref{pert-2}, and \eqref{pert-5}, which can be written in a matrix form as
  \begin{equation} \label{stability-equation}
 {\cal M}\left( {\begin{array}{*{20}c}
   C_B  \\
   C_Q  \\
   C_{Q_2} \\
 \end{array} } \right) \equiv \left[ {\begin{array}{*{20}c}
   {\mu} & {2B} & {0 }   \\
   { 0} & {\mu} & {-1 }  \\
     {0  } & {\frac{1}{44\bar\gamma} \left( 2B^3 -96\bar\gamma \right) } & {  \mu+3 }  \\
 \end{array} } \right]  \left( {\begin{array}{*{20}c}
   C_B  \\
   C_Q  \\
   C_{Q_2} \\
 \end{array} } \right) = 0.
\end{equation}
Mathematically, this homogeneous system admits non-trivial solutions $C_i \neq 0$ with $i=B,~Q,$ and $Q_2$ if and only if
\begin{equation}
\det {\cal M} =0,
\end{equation}
which can be calculated to be a cubic equation of $\mu$ given by 
\begin{equation} \label{quadratic-equation}
\mu  \left(11\mu^2+33\mu -18  \right) =0,
\end{equation}
thanks to the value of the de Sitter fixed point, $B =\left(12\bar \gamma \right)^{1/3}$. As a result,  three values of $\mu$ can be easily solved from Eq. \eqref{quadratic-equation} to be
\begin{equation}
\mu_1 =0, \quad \mu_{2} = -\frac{3}{22} \left(11 +   \sqrt{209} \right) <0 ,\quad \mu_{3} = -\frac{3}{22} \left(11 -   \sqrt{209} \right) > 0.
\end{equation}
These roots are indeed consistent with that shown in Eq. (29) of the original paper of the EGZ gravity \cite{CamposDelgado:2024jst}. It is noted that the authors of Ref. \cite{CamposDelgado:2024jst} have considered only the field equation \eqref{equation-1-FLRW} for their stability analysis. The existence of the positive root $\mu_3$ clearly indicates that the de Sitter fixed point is indeed unstable  since all exponential perturbations will blow up as the dynamical time $\tau$ becomes large. Hence, the corresponding de Sitter solution obtained in the previous section is unstable, too. Furthermore, the values of $\mu_1$, $\mu_2$, and $\mu_3$ do not depend on the values of $\zeta$ and field parameters, meaning that both de Sitter expanding and inflationary solutions of the EGZ gravity are always unstable. This result is indeed consistent with that obtained in the original paper of the EGZ gravity \cite{CamposDelgado:2024jst}.  In addition to the instability of the de Sitter solution, we are able to show, by using numerical calculations, that the de Sitter fixed point is indeed a repeller of the dynamical system.  This behavior can be seen from Fig. \ref{fig1}.  In particular, this figure displays three different trajectories starting with three different initial conditions and two distinct points, the red and black ones, corresponding to two mentioned fixed points of the dynamical system.  It is noted that the number of trajectories has been chosen to be three just for simplicity. Of course, one can plot more trajectories to have a high-resolution visualization. The red point corresponds to one fixed point mentioned above, whose coordinates in the phase space are $(B,Q,Q_2) \simeq (0,0.0796,0.0127)$ for $\gamma=10^{-6}$ and $M_p=1$; while the black point corresponds to the de Sitter fixed point, whose coordinates in the phase space are $(B,Q,Q_2) \simeq (0.023,0,0)$ also for $\gamma=10^{-6}$ and $M_p=1$. As the dynamical time $\tau$ increases, all three trajectories tend to converge to the red point rather than the black point. This behavior clearly indicates that the red point is an attractor and the black point is a repeller of the dynamical system \cite{Bahamonde:2017ize}. In other words, the de Sitter spacetime seems not to be a final background of the universe, both in the early and late-times, within the context of the EGZ gravity.

The instability property of the de Sitter solution indicates that the EGZ gravity seems to be more compatible with an inflationary phase of the early universe rather than an accelerated expansion of the late-time universe, according to the discussions in Refs.  \cite{Elizalde:2014xva,Pozdeeva:2019agu,Vernov:2021hxo}. This claim is due to the fact that  a stable and attractive de Sitter inflationary solution would lead to the so-called eternal inflation and therefore a multiverse scenario \cite{Guth:2007ng}. In this case, the so-called graceful exit mechanism is supposed to happen in order to ensure a smooth connection between an early inflationary phase and a late-time expanding FLRW phase \cite{Brustein:1994kw}. Unfortunately, figuring out such a mechanism is not always straightforward. Hence, any gravity model admitting an unstable de Sitter inflationary solution would be more realistic since it can easily overcome the eternal inflation issue.  Furthermore, if we still use the EGZ gravity for the late-time universe, the existence of the so-called dark energy \cite{Copeland:2006wr,Nojiri:2005sx,Nojiri:2006je,Carter:2005fu} may be necessary  in order to achieve a stable de Sitter expanding solution within this gravity. It is worth noting that the appearance of higher-order curvature corrections may play an important role in removing future singularities such as the Big Rip singularity, which could caused by the dark energy \cite{Nojiri:2005sx}, as indicated in the interesting review paper on the dynamics of dark energy  \cite{Copeland:2006wr}. Remarkably, the dynamical system method has been used to investigate the future singularities \cite{Nojiri:2005sx}.
\section{Conclusions}\label{final}
We have investigated the stability of de Sitter solution in the EGZ gravity proposed recently in Ref. \cite{CamposDelgado:2024jst} by using the dynamical system method. As a result,  the value of the obtained de Sitter solution is identical to one derived in Ref. \cite{CamposDelgado:2024jst}. More importantly, the de Sitter solution is apparently unstable against field perturbations, similar to the conclusion of the original paper of the EGZ gravity \cite{CamposDelgado:2024jst}. Furthermore, it has been numerically confirmed to be a repeller as displayed in Fig. \ref{fig1}. It should be noted that these results have been worked out despite the fact that one of the field equations derived by using the EL equations is not identical to one derived in the original paper of the EGZ gravity \cite{CamposDelgado:2024jst}. The instability property of the de Sitter solution implies that the EGZ gravity would be more relevant to the inflationary phase of the early universe rather than the accelerated expansion of the late-time universe. Since the main goal of the present paper is solely  the stability analysis of de Sitter solution, further cosmological implications of the EGZ gravity will be our future studies. Currently, it is hoped that the present paper could be useful to studies of stability analysis of fourth-order gravities, e.g., the Einsteinian cubic (ECG) gravity  \cite{Bueno:2016xff,Arciniega:2018fxj} and the Starobinsky-Grisaru-Zanon (SGZ) gravity \cite{Toyama:2024ugg}. In particular, the  the stability of de Sitter solutions within the context of the ECG as well as SGZ gravities has remained unclear. One could therefore extend our current analysis to these models to seek answers. We would like to emphasize again that examining the existence and/or stability of de Sitter solution in higher-order gravity models is one of the most important initial checks before investigating further their cosmological viability and consequences \cite{Elizalde:2014xva,Pozdeeva:2019agu,Vernov:2021hxo,Faraoni:2004dn,Faraoni:2005vk,Toporensky:2006kc,Kamenshchik:2024kay}. For example, if a higher-order gravity always admits an unstable de Sitter solution then the existence of the so-called dark energy \cite{Copeland:2006wr,Nojiri:2005sx,Nojiri:2006je,Carter:2005fu} may be necessary in order to achieve a stable de Sitter expanding solution within this gravity. Moreover, the inclusion of higher-order curvature corrections may play an important role in removing future singularities such as the  Big Rip singularity, which could be caused by the dark energy \cite{Copeland:2006wr,Nojiri:2005sx}. Besides the stability issue, our paper clearly demonstrates the usefulness and effectiveness of the EL equations in defining complicated field equations of higher-order gravities or at least in cross-checking those defined by the  Einstein (tensorial) field equations. It has been shown that one of field equations defined incorrectly in Ref. \cite{CamposDelgado:2024jst} has been corrected by using the EL equations. Further consequences due to this error will therefore be avoided. 
\begin{acknowledgments}
This study is funded by the Vietnam National Foundation for Science and Technology Development (NAFOSTED) under grant number 103.01-2023.50. The author would like to thank Prof. Phung V. Dong very much for his support.
\end{acknowledgments}

\end{document}